# The Long-Run Impact of Electoral Violence on Health and Human Capital in Kenya

Roxana Gutiérrez-Romero♣

**Abstract**

This paper examines the long-term effects of prenatal, childhood, and teen exposure to electoral violence on health and human capital. Furthermore, it investigates whether these effects are passed down to future generations. We exploit the temporal and spatial variation of electoral violence in Kenya between 1992 and 2013 in conjunction with a nationally representative survey to identify people exposed to such violence. Using coarsened matching, we find that exposure to electoral violence between prenatal and the age of sixteen reduces adult height. Previous research has demonstrated that protracted, large-scale armed conflicts can pass down stunting effects to descendants. In line with these studies, we find that the low-scale but recurrent electoral violence in Kenya has affected the height-for-age of children whose parents were exposed to such violence during their growing years. Only boys exhibit this intergenerational effect, possibly due to their increased susceptibility to malnutrition and stunting in Sub-Saharan Africa. In contrast to previous research on large-scale conflicts, childhood exposure to electoral violence has no long-term effect on educational attainment or household consumption per capita. Most electoral violence in Kenya has occurred during school breaks, which may have mitigated its long-term effects on human capital and earning capacity.

**JEL codes**: I15, I12, O15, O55

**Keywords:** electoral violence, long-term effects, stunting, educational attainment, Kenya, Africa.

♣ Queen Mary University of London, School of Business and Management, London, UK. r.gutierrez@qmul.ac.uk. I am grateful for the hospitality of St. Antony's College and the Department of Economics at the University of Oxford, where I spent my sabbatical leave while working on this article. I also thank the staff members of Infotrak Kenya, especially Raphael Mulwa and Jimmy Kiprono, for their assistance with the survey. This work was supported by the UK Department for International Development. I also acknowledge financial support from the Spanish Ministry of Science and Innovation (ECO2013-46516-C4-1-R) and the Generalitat of Catalunya (2014 SGR 1279).

# 1. Introduction

Electoral violence is a pervasive and recurrent problem in many parts of the developing world. In Africa, for example, nearly 60% of elections have been marred by violence, far exceeding the global average of 19% (Blattman and Miguel, 2010; Burchard, 2015). During outbreaks of electoral violence, many families experience food shortages, are relocated against their will, or are forced to remain indoors for several weeks to avoid being victimized (Bekoe, 2012). Typically, electoral violence is less deadly and shorter in duration than wars and other major armed conflicts (Goldsmith, 2015). Still, this type of political violence causes a wide range of immediate economic and humanitarian costs. Households affected by the turmoil may be forced to take costly measures to smooth their consumption, such as engaging in risky activities to generate more income, selling assets, temporarily reducing food intake, and even withdrawing children from school (Dupas and Robinson, 2012; Verwimp et al., 2019). Given the frequency with which electoral violence occurs, particularly in low-income countries, it is critical to gain a better understanding of the short- and long-term effects on well-being. It is especially important to learn about these effects for understudied vulnerable groups in such conflicts, such as children and adolescents.

Extensive research on major armed conflicts has provided important insights into the long-term effects of household consumption shocks that result in malnutrition in young children. Studies in Burundi, Côte d'Ivoire, Ethiopia, Rwanda, and Zimbabwe, among others, have found negative effects on children's height and education (Agüero and Deolalikar, 2017; Akresh et al., 2012b; Alderman et al., 2006; Bundervoet et al., 2009; Minoiu and Shemyakina, 2014). Other, more widespread, and deadlier civil conflicts have been found to pass on nutritional shocks to future generations. This was the case during Nigeria's three-year civil war, which resulted in 100,000 military casualties and up to three million civilian deaths due to starvation. This conflict lowered the stature of both children exposed to it and their descendants (Akresh et al., 2021). Despite a large body of research in this field, there is a significant research gap regarding the long-term effects of low-casualty intensity, short-lived electoral violence on health and human capital. The potential long-term effects of prenatal, childhood, and adolescent exposure to electoral violence are particularly poorly understood. In addition, no research has been conducted to determine if the consumption and income shocks that this type of conflict can cause can be transmitted to the descendants of those who were exposed to electoral violence.

To bridge this research gap, this paper examines two research questions. First, the paper evaluates whether prenatal, childhood, and adolescent exposure to electoral violence in Kenya has had any long-term impact on health and human capital. In particular, we examine three long-term outcomes: adult height, educational attainment (highest level of education completed), and household consumption per capita. Second, the paper also investigates whether these long-term effects on height and educational attainment are passed down to the children of people who were exposed to electoral violence when they were growing up. We conducted a nationally representative survey designed specifically to test our research questions. This survey asked adult respondents whether they had been exposed to electoral violence, how, and when. The survey also collected rich socio-economic and anthropometric information about the respondents, their parents, and their biological children.

We focus on Kenya, a country that since the re-introduction of multiparty elections in the 1990s has experienced electoral violence in every single election, like many other new democracies in Sub-Saharan Africa. The violence has resulted in a lower casualty rate and shorter duration than other large-scale armed conflicts such as civil wars. For instance, in Kenya, between 1991 and 1998, 2,000 people were killed and 400,000 others were forcibly displaced as a result of electoral violence (Human Rights Watch 2002, p.21). In just three months, the unrest that followed the announcement of the 2007 presidential election resulted in the deaths of 1,133 people and the displacement of over a quarter of a million others (CIPEV 2008). The next election of 2013, despite being widely regarded as peaceful, still claimed the lives of 500 people (Human Rights Watch, 2013). This recurring nature of electoral violence displays variation in which areas have been affected over time. This variation in exposure to violence across time and space allows us to evaluate the long-term effects of violence across generations. In other words, we evaluate the effects on those directly exposed to electoral violence during prenatal, childhood, and adolescence, as well as on their descendants.

Our intuition is that households affected by electoral violence will experience significant temporary and unanticipated fluctuations in income and consumption. These shocks are likely to force affected families to reduce their nutrient-rich food and calorie intake, negatively impacting the height of children and teens, and that of their descendants. Because malnutrition can impair cognitive performance, nutritional shocks may also impact academic performance and potentially long-term earning capacity (Berkman et al., 2002). In addition, as other large-scale and protracted conflicts have demonstrated, the nutritional and stunting effects of conflict can be passed on to future generations (Akresh et al., 2021), an important aspect we investigate for our case study.

We determine which of our survey respondents were victims of electoral violence during their growing years (from prenatal to adolescence) using two alternative measures. First, based on the respondents' birth year and whether claimed that they, their households, or their villages had experienced electoral violence in previous elections, and if so, which election. Separately, we identify the victims of the electoral violence of 2007–08 based on the adult respondents' birth year and their reported district of residence during that disputed election. Using information from the Commission of Inquiry on Post-Election Violence, we determine which of these districts experienced electoral violence-related injuries or fatalities in 2007–08. Then, to evaluate the impact of electoral violence, we compare the height of adults who, during their growing years (from gestation up to 16 years old), were living in households or villages affected by electoral violence to the height of similar respondents who were not affected by electoral violence. Some victims of electoral violence may have characteristics that increase their likelihood of experiencing violence, such as living in poverty, and cause them to have reduced height. Thus, to identify the impacts of electoral violence, we use the coarsened matching estimator, as proposed by Iacus, King, and Porro (2008). This nonparametric matching method helps reduce some or all potentially confounding factors by reducing significant differences in characteristics between people affected and unaffected by electoral violence. In addition, we use a variety of strategies to demonstrate that our findings are not due to misspecification, endogenous selection, or other factors, such as a decline in agricultural production due to extreme weather. We also present placebo tests in which we find no effect on the height or human capital of those affected by electoral violence as adults, nor on their descendants.

This paper makes three contributions. Previous research on large-scale civil wars has shown that young children exposed to the conflict have shorter stature as a result of associated consumption and nutritional shocks (Akresh et al., 2021). Recent studies have also shown that children exposed to Zimbabwe's five-year political violence, which displaced many farm owners and workers between 2000 and 2005, were stunted as a result (Shemyakina, 2021). Our first contribution to the literature is to show that even small-scale electoral violence, with fewer casualties and lasting only a few weeks can also have lasting health effects. Specifically, we show adults in Kenya who were exposed to electoral violence during their prenatal, childhood, or adolescence are shorter in height than similar adults who were not exposed to such violence during their growing years.

Our second contribution focuses on the intergenerational effects of low-scale, recurring electoral violence, such as that seen in Kenya. We show that this type of violence has also affected the height-for-age of children whose parents were victims of such violence during their growing age. Remarkably, this intergenerational effect is only found in boys. Two alternative tests back up this result. For example, we examine the relationship between parental victimization of electoral violence at a young age and the offspring's sex on height-for-age z-scores. We find that sons (newborn to 18 years old) of parents affected by electoral violence are shorter than their daughters, compared to the height difference between sons and daughters of non-affected parents. Furthermore, boys whose parents experienced electoral violence during their childhood or adolescence are also shorter than boys whose parents were unaffected by such violence. Our findings suggest that since boys in Kenya, as in the rest of Sub-Saharan Africa, have an increased risk of malnutrition and stunting, a significant nutritional shock experienced by their parents can be passed down and have a greater impact on them (Keino et al., 2014; Wamani et al., 2007).

The third contribution is to show that neither the educational attainment of adults exposed to electoral violence during their growing years nor that of their descendants has been impacted by the type of electoral violence examined. A limitation of our study is that we rely on self-reported levels of education successfully completed and lack official records of our respondents' test scores or attendance, which may have been affected. As an alternative test for long-term effects on human capital formation, we also examine household consumption per capita and find no effect. Recent research on the impact of major armed conflicts on educational attainment in Africa and elsewhere contradicts our findings (Akbulut-Yuksel, 2014; Akresh et al., 2021). Our lack of a significant impact on human capital could be attributed to several factors, including Kenyan electoral violence being on a small scale and concentrated during school holidays, which may have limited its effects on human capital. Nonetheless, the paper demonstrates that electoral violence can leave a destructive legacy over the long term. These effects have a disproportionate impact on both children's height (which is a good indicator of their health during their growing years) and even that of their descendants. The results of this study are important as they can provide evidence that low-scale electoral violence can have lasting impacts on children and future generations, highlighting the need for measures to prevent and mitigate such violence.

The remainder of the paper is as follows. The next section explains the causes of the repeated electoral violence in Kenya. The literature on the effects of electoral violence on health and education is then discussed. We then show the data, discuss the identification strategy, present the results, robustness checks, followed by a discussion of the results.

## 2. Setting

Kenya is a good example of how electoral violence can recurrently lead to forced displacement, food insecurity, and turmoil. Following its independence from Great Britain in 1963, Kenya became a one-party state (first *de facto* and later *de jure*), with the Kenya African National Union (KANU) party winning all subsequent elections in 1969, 1974, 1979, 1983, and 1988. Following domestic and international pressure, Kenya held its first multiparty elections in 1992 and 1997. Both elections were won by KANU amid unprecedented electoral violence. To a large extent, the violence was state-sponsored.

The KANU's government, led by Daniel Arap Moi, sought to retain power by exploiting land grievances that long predated the reintroduction of multiparty elections. Just ahead of the elections, Moi's coethnics in the Rift Valley were encouraged to reclaim their ancestral land that the first president of Kenya, Jomo Kenyatta, had unfairly redistributed or 'stolen' decades earlier and given to non-indigenous people, primarily the Kikuyu, Kenyatta's coethnics. The ethnically charged rhetoric posed a genuine threat. Unexpectedly, state-trained militias launched a campaign to evict non-indigenous groups from the Rift.[1]

The impunity of state actions drove ordinary citizens to join in ethnic cleansing raids, killing, burning houses of non-indigenous groups, and evicting them from their land, expecting it as a reward for their loyalty to Moi's regime. These militia members were given clear incentives, $25 dollars per killing and up to $125 per house burned to ashes (Mulli, 1999). These raids affected the Rift Valley and borderland areas of Nyanza and Western Provinces through 1992–1998. By 1997, ethnic cleansing had erupted in the Coast province as land grievances drove locals to kill and evict non-indigenous ethnic groups. Boone (2011) shows electoral violence in all areas where KANU had established land settlement schemes or sold land to non-indigenous groups decades earlier. The 1990s electoral violence spared the Trust territory, 75% of Kenya, where much of the land is held under customary rights by ethnic communities and land rights were not politicized.

Since the reintroduction of multiparty elections, elections have become a vehicle for the elite and opposition groups to reclaim ancestral land rights. Every election, these land grievances resurface. As a result of the violence, several small and medium-sized farms have been impacted, particularly in the Rift Valley, one of the country's most agriculturally productive areas. The region primarily serves local markets and the country's top exports, such as avocado, coffee, and tea, as well as the large flower exporting industry, which also provides jobs to workers in the region (Debonne et al., 2021; Macchiavello and Morjaria, 2015).

---

[1] These grievances date back to the colonial era, when British colonists seized roughly 20% of Kenya's prime agricultural land in the Rift Valley, Western, and Nyanza. Upon Kenya's independence, Britain returned about half of the seized land. However, the communities from which the land was taken did not receive it back. Kenyatta instead sold the land on the basis of 'willing seller, willing buyer', disregarding the communities' customary collective land rights (Human Rights Watch, 2008). This new land tenure system provided the state opportunities to manipulate land allocation and offer land rights in exchange for patronage (Boone, 2011).

Between 1991 and 1998, electoral violence killed 2,000 people and forced more than 400,000 to leave their homes. The violence continued through the by-elections for the parliament in 2001 and into the campaign for the presidency in 2002 (Human Rights Watch, 2002). Nonetheless, the 2002 election was a genuine contest between parties, resulting in a peaceful transfer of power (Collier et al., 2010). Promising to resolve the land disputes and introduce constitutional reforms, Mwai Kibaki beat KANU's presidential candidate, Uhuru Kenyatta. The next presidential election of 2007 was expected to pass off equally smoothly, but surprisingly, Kibaki was announced as the winner of the 2007 election despite the early count of the votes. Within minutes of Kibaki's reelection announcement, youth gangs in the Rift, Nairobi, Coast, and Western provinces began torching homes, murdering, raping, and looting. In February 2008, the two leading candidates for president, Mwai Kibaki and Raila Odinga, signed a power-sharing agreement. By that time, 1,133 people had been killed and over 250,000 had been displaced.

Nearly 30% of households interviewed in a nationally representative survey conducted shortly after the 2007–08 electoral violence reported being directly affected by violence in terms of personal injury, displacement, or economic loss (Gutiérrez-Romero, 2014). The closure of markets, especially near the epicenter of the violence, caused food shortages and inflation (Glauser, 2008). During the period of violence, firms located in the Rift and Western provinces reported an average of 50% of their workforce was missing. In contrast, firms located in other regions experienced no significant effects (Macchiavello and Morjaria, 2015). Nearly 20% of the country's maize crops, the main staple food, were not harvested due to political unrest. Additionally, stores and markets were closed for several weeks. The closures resulted in significant losses of already-harvested crops, perishables such as milk and vegetables, and income-earning opportunities in the agricultural sector, the country's most important economic driver, as is the case in much of Sub-Saharan Africa (FEWS, 2008).

In Bumala, one of the districts severely affected by violence, Dupas and Robinson (2012) find that the price of essential foods such as milk, sugar, cooking oil, and other necessities increased by up to 30% in the weeks following the electoral dispute. Even after the power agreement that ended the violence, the inflation and shortage of food continued for months. Although the 2013 election that followed avoided a major political crisis, it was again marred by electoral violence that claimed 500 lives (Human Rights Watch, 2013).

# 3. Consequences of electoral violence

## 3.1 Food supply

Throughout history, large-scale armed conflicts have disrupted food systems, resulting in hunger vulnerability, poverty, and malnutrition that can persist long after civil unrest has subsided (Messer, 1998). In conflict-prone regions such as Asia and Africa, governments and rebels frequently use food control as a strategic weapon to starve opponents and recruit potential followers (Messer, 1990). During the 1970s and 1980s, for instance, fourteen out of sixteen wars in the developing world significantly reduced food production (Stewart, 1993). In 2004, more than 45 million people who were experiencing or recovering from conflict required food or humanitarian aid, with the majority, 80%, residing in Sub-Saharan Africa (Messer and Cohen, 2014).

Less is known about the effects of electoral violence on food supply and food insecurity. If rebels or incumbent governments target specific groups during elections and strip them of their land and other assets necessary to invest in their future food supply, this could have long-term effects on income and health. Similarly, electoral violence can indirectly affect the food supply by disrupting transportation and trade networks. As was the case in Kenya, disruptions in trade networks can hinder trade with nearby food markets and prevent workers from returning to the field to restore food production (Macchiavello and Morjaria, 2015). Food-producing regions are paradoxically more susceptible to 'food war tactics' because it is less difficult to seize or destroy food stocks as a strategic weapon (Gutiérrez-Romero, 2022; Koren and Bagozzi, 2017). Additionally, electoral cycles may increase the likelihood of land disputes, resulting in a concentration of food insecurity in particular regions (Boone, 2011; Klaus and Mitchell, 2015). In these land disputes, elites frequently use grievances to incite violence, leading to ethnic divisions, forced migration, and food insecurity. Vulnerable groups may experience long-term losses in their production factors, making it difficult to recover their economic and food production.

Temporary migration to cities, which is common in Sub-Saharan Africa, has the potential to exacerbate the disruption caused by electoral violence. Rural residents' ability to send money home may be hampered by violence, reducing their income and ability to withstand sudden drops in income. If political barriers prevent victims of electoral violence from receiving humanitarian assistance, they may face long periods of food insecurity and significant dietary changes (Messer, 1998). Consequently, if a household suffers a significant income shock, all affected family members, including children, may be required to reduce their nutrient and caloric intake. This could explain why women were found in Kenya doing dangerous or extra work to maintain their living standards following the 2007–08 electoral violence (Dupas and Robinson, 2012).

*3.2 Height*

Extensive research on large-scale and protracted civil wars has confirmed that food supply disruptions caused by violence can have long-term effects on the height of children and adolescents exposed to the conflict (Akbulut-Yuksel, 2014; Akresh et al., 2012a; Balalian et al., 2017; Minoiu and Shemyakina, 2014). Armed conflict exacerbates malnutrition, which affects hundreds of millions of pregnant women and children in developing countries (Müller and Krawinkel, 2005). According to previous research, children who were malnourished as a result of widespread armed conflict may experience stunted growth as adults. That is, these children who have been exposed to conflict may not catch up during their adolescent second growth spurt (Bradley and Corwyn, 2002). Nutritional shocks can have long-term consequences even in utero. As evidenced by low birth weight and other elevated health risks, a woman's nutrition during pregnancy has an impact on the physiological and neurological development of the fetus (Alderman and Behrman, 2006).

Adolescents exposed to armed conflict may also suffer from severe nutritional deficiencies, resulting in stunting. For instance, Akresh et al. (2021) find that children and adolescents who experienced food insecurity during Nigeria's civil war between 1967 and 1970 had stunted height as adults. Women and adolescents were most affected. These researchers also found evidence of intergenerational effects, as children born to mothers who had been exposed to the war were shorter in height.

These results suggest that large-scale conflicts, such as civil wars, can affect the height of the second generation. However, little is known about whether smaller-scale violence, such as that seen during Kenyan elections, can also have these health consequences. We argue that this is a possibility here. Human height is inherited from both parents, and there is substantial evidence that short adult height results from malnutrition's intergenerational effects (Khatun et al., 2019). The intergenerational effects of height on linear growth have also been demonstrated in animal models. Consistent with the secular anthropometric height trends observed in Europe and North America, these animal studies show that it takes multiple generations to eliminate the adverse height effects of malnutrition (Martorell and Zongrone, 2012). These long-term trends refer to population differences caused primarily by differences in birth dates. Although epigenetics and intergenerational poverty transmission are factors in adult stunting, there is a strong intergenerational link between short maternal height and an increased risk of infant mortality and stunting. That is the case even when socioeconomic factors are controlled for (Khatun et al., 2018, 2019).

Kenya, our case study, has experienced electoral violence in every election since the 1990s. This violence lasts for a few weeks to months and is less deadly in comparison to civil wars.

Nonetheless, the repeated disruptions to the food chain caused by electoral violence may have long-term and significant consequences for the nutrition of children and adolescents exposed to violence. Based on this discussion, we hypothesize that children and adolescents exposed to electoral violence in Kenya will grow shorter than comparable youths who were not exposed to such violence. Moreover, if the nutritional shocks caused by electoral violence were severe, the effects of stunting may have been passed down to future generations.

*3.3 Educational attainment*

Multiple studies have found that stunted growth in childhood is a strong indicator of poorer cognitive performance and worse educational outcomes (Behrman and Rosenzweig, 2004; Case and Paxson, 2008). For instance, Alderman, Hoddinott, and Kinsey (2006) find that the civil war in Zimbabwe decreased the child height-for-age z-score by 0.5 points. This reduction is very similar to the 0.6-point decrease in z-scores experienced by the severe drought that followed from 1982 to 1984. These stunted children remain stunted as adults and have poorer educational outcomes. Similarly, Akresh et al. (2021) find that the Nigerian civil war also had intergenerational impacts on education. In other words, women who had been exposed to civil war were more likely to have lower educational attainment. Furthermore, mothers who were exposed to the civil war had a negative impact on their children's education.

Aside from nutritional shocks, it is well known that schooling may be disrupted as conflicts unfold (Miller et al., 2008). Some children may be required to leave school to assist their families in dealing with the effects of consumption shocks or forcible displacement. Despite the extensive literature on how civil conflict affects health and education outcomes, it is unclear whether electoral violence, which is typically small-scale and brief, can have similar effects, and whether these effects last into adulthood or even second generations. This is even less obvious in Kenya, where election-related violence has primarily occurred during school holidays. Based on the specifics of our case study, we hypothesize that the small-scale, brief, and school-holiday-centered electoral violence will not affect the educational attainment of those who experience it or their descendants.

**4. Survey**

To estimate the long-term impact of electoral violence, we conducted a nationally representative survey in Kenya. The survey interviewed 1,210 adults of voting age across 80 of the 290 constituencies in the country in December 2013. The random sample was stratified by province, urban, and rural levels, using the same sampling techniques as the Afrobarometer, a widely used survey in the region.[2] Our survey interviews were conducted in person. All adult respondents were questioned about their socioeconomic status, election-related victimization, education, employment, and health, including during their childhood and adolescence. Table A.1 in the appendix provides a summary of the characteristics of the adult respondents interviewed. Additionally, these respondents were asked about their parents and biological children.

*4.1 Identifying respondents exposed to electoral violence*

Table 1 displays the birth cohorts of generations born during the five most recent elections at the time of the survey (1992, 1997, 2002, 2007, and 2013) who may have been exposed to electoral violence during their prenatal, childhood, or adolescence. In total, 370 of the 1,210 adult respondents interviewed (all of whom were over the age of 37 at the time of our survey) were born before the cohorts depicted in Table 1. Although they may have been affected by electoral violence, if so, it occurred after they were adults and not during their prenatal, childhood, or adolescence. Still, we examine this subsample, but only as a robustness check to see if electoral violence, for example, did not affect their height. Instead, we will examine the remaining 840 of the 1,210 adult respondents who were born during the cohorts highlighted in Table 1. We use this subsample of respondents (aged 18 to 37 at the time of the interview) to test the potential long-run effect of being exposed to electoral violence while in utero, childhood, or adolescence.

**Table 1** Birth cohorts exposed to electoral violence.

| Year of election | | 1992 | 1997 | 2002 | 2007-08 | 2013 |
|---|---|---|---|---|---|---|
| Birth cohort at the time of the election | Aged 13 to 16 | 1976-1980 | 1981-1988 | 1986-1990 | 1991-1995 | 1997-2001 |
| | 7 to 12 | 1980-1983 | 1985-1991 | 1990-1996 | 1995-2001 | 2001-2007 |
| | 4 to 6 | 1983-1989 | 1991-1994 | 1996-1999 | 2001-2004 | 2007-2010 |
| | in utero to 3 | 1989-1992 | 1994-1997 | 1999-2002 | 2004-2008 | 2010-2013 |

The birth cohorts highlighted in grey were adults of voting age (18+) at the time of our survey.

We use two proxies to identify respondents who were exposed to electoral violence as adults or earlier in their lives. The first proxy is based on respondents' birth year, which we compare to the birth cohorts shown in Table 1, as well as whether respondents reported being exposed to electoral violence directly (themselves or their households) or indirectly (their village). We asked for this information for each of the previous five multiparty elections. Table 2 shows the six types of questions we used to identify respondents who had been 'exposed to electoral violence' in the previous five elections. Two of these questions ask if respondents and their family members have been impacted by electoral violence in each of the following ways: damage to property, destruction of home, being forced to leave their home, destruction or closure of a business, loss of a job, personal injury, and land disputes. We also asked three separate questions about whether respondents' households saw a decrease in income and consumption, as well as whether they were displaced from their homes or land as a direct result of electoral violence. Lastly, we asked respondents if their village of residence experienced electoral violence on a regular basis, and if so, in which election. As Table 2 shows, 20% of the 1,210 adult respondents interviewed reported being affected by the electoral violence of 2007–08. In line with our expectations, this disputed election had the highest rate of victims in terms of personal impact, family members, and the villages affected. Taking all elections into account, 19% of all adult respondents were exposed to electoral violence during their early life, including in utero, childhood, or adolescence.

---

[2] The Afrobarometer's sampling procedure was followed (https://www.afrobarometer.org/surveys-and-methods/sampling). We chose 80 constituencies, primary sampling units, and households at random after calculating the sample size to be nationally representative.

**Table 2** Descriptive statistics on electoral violence exposure.

| Year of election | | 1992 | 1997 | 2002 | 2007 | 2013 |
|---|---|---|---|---|---|---|
| Have you personally ever been affected by the outbreaks of electoral violence in terms of damage to personal property, destruction of your home, being forced to leave your home, destruction or closure of a business, loss of a job, personal injury, or land disputes? | | | | | | |
| | Mean | 0.019 | 0.012 | 0.014 | 0.206 | 0.088 |
| | St. Dev. | 0.137 | 0.111 | 0.118 | 0.404 | 0.283 |
| Have you ever been displaced, leaving your home or land in response to incidents of electoral violence? | | | | | | |
| | Mean | 0.010 | 0.004 | 0.007 | 0.053 | 0.007 |
| | St. Dev. | 0.099 | 0.064 | 0.081 | 0.224 | 0.081 |
| Because of electoral violence, was your household's income reduced? | | | | | | |
| | Mean | 0.050 | 0.056 | 0.079 | 0.367 | 0.240 |
| | St. Dev. | 0.217 | 0.230 | 0.270 | 0.482 | 0.427 |
| Because of electoral violence, was your household consumption reduced? | | | | | | |
| | Mean | 0.057 | 0.055 | 0.087 | 0.398 | 0.231 |
| | St. Dev. | 0.232 | 0.229 | 0.282 | 0.490 | 0.421 |
| Have any of your family members ever been affected by the outbreaks of electoral violence in terms of damage to their personal property, destruction of their home, being forced to leave their home, destruction or closure of a business, loss of a job, personal injury, or land disputes? | | | | | | |
| | Mean | 0.037 | 0.015 | 0.019 | 0.150 | 0.058 |
| | St. Dev. | 0.189 | 0.121 | 0.137 | 0.357 | 0.234 |
| Did electoral violence occur frequently in your village in any of the following elections? | | | | | | |
| | Mean | 0.054 | 0.043 | 0.037 | 0.200 | 0.064 |
| | St. Dev. | 0.226 | 0.203 | 0.189 | 0.400 | 0.244 |

We also use a second alternative proxy to identify respondents who have experienced electoral violence. This is based on the respondents' birth year and the district in which they reported living during the electoral violence of 2007–08. Then, we use the Commission of Inquiry on Post-Election

Violence, (CIPEV 2008), data to determine which districts experienced injuries or fatalities as a direct result of the 2007–08 electoral violence. This episode is analyzed in depth as it is by far the most violent in Kenyan history. This commission identified 1,133 deaths and 3,561 injuries as a result of electoral violence in 39 of the 70 official districts. Specifically, injuries were identified in 32 districts, and casualties in 28 districts affecting five provinces (Nyanza, Western, Rift Valley, Mombasa, and Nairobi). In comparison to previous episodes of electoral violence, which primarily affected a small number of districts in the Rift Valley, Western, and coastal region, the 2007–08 election violence was more widespread, affecting both rural and urban areas (CIPEV 2008).

*4.2 Measuring height, human capital, and household consumption per capita*

We continue by describing the relevant outcomes collected in the survey to evaluate the impact of electoral violence exposure. Interviewers measured the heights of all adult respondents (aged 18 to 60) and their biological children (aged newborn to 18). The interviewers measured the height of non-ambulatory infants and toddlers with the help of the child's parents. Children were positioned horizontally by laying them on a flat surface with their heads and feet against a vertical surface and applying gentle pressure to their knees to straighten their legs.

We asked respondents what their 'highest level of education they had completed' was to assess the impact on human capital. We also inquired about the educational attainment of each respondent's biological children aged 18 and under. We asked about the educational attainment of the respondents' father when he was of the respondent's age, or when he died if he died before this age, to gain a better understanding of the potential household wealth when the respondents were growing. We requested this information about the mother if the respondent stated that he had never met his father.

The survey asked respondents 'how much money does your household spend per month' to assess their contemporaneous monthly household consumption. We can estimate the respondent's household consumption per capita at the time of the interview because we also asked, 'how many people live in your household, including children?'.

To get a sense of the respondent's current level of wealth, we also create a standardized wealth index that ranges from 0 to 1 based on whether the respondent claimed to possess each of the following fifteen durable assets: house, land, livestock, oven, refrigerator, washing machine, computer, phone, mobile phone, book, radio, television, bicycle, and motorcycle. Based on the following question, we created another standardized wealth index for the wealth of the respondent's father: 'Did your father's household own any of the following durable assets when he was your age, or when he died if before this age?' We requested the same list of assets as before, except for mobile phone ownership. We inquired about their mother's assets from respondents who claimed not to have met their father.

*4.3 Descriptive statistics of victims of electoral violence and non-victims*

Table A.2 contains additional information about adult respondents who have been exposed to electoral violence since 1992, whether during their prenatal, childhood, or adolescence. Most of these now-adults (75%) agreed to have their height measured. In terms of average height, respondents who were exposed to electoral violence in their early years are nearly two centimeters shorter than those who were not. At the time of the survey, victims and non-victims of electoral violence had comparable educational attainment, monthly household consumption per capita, and wealth index.

Table A.3 provides descriptive statistics on respondents' biological children (ranging in age from newborns to 18 years). We use height-for-age z-scores to make meaningful height comparisons between children of different ages and sexes. Z-scores, which represent the number of standard deviations below or above a reference mean for children and adolescents of a given age and gender, are a good indicator of growth retardation and malnutrition. The WHO Reference Charts from 2007 serve as the population of reference. Children with a height-for-age z-score less than -2 (more than two standard deviations) from the mean of healthy children are considered stunted and malnourished. According to our survey, boys are more likely than girls to be stunted.

## 5. Identification strategy

We use two methods to assess the long-term impact of prenatal, childhood, and adolescent exposure to electoral violence. To begin, we use Ordinary Least Squares (OLS) regressions, as illustrated in Eq. (1). Then, to address potential endogeneity concerns, we refine our regressions further using coarsened matching, as described further below.

$$outcome_{iay} = \beta \: Victim \: growing \: age_{iy} + \delta X_{iay} + \alpha_{iy} + \varphi_{ie} + \mu_{ia} + \varepsilon_i \quad (1)$$

The dependent variable, $outcome_{iay}$, represents the long-term outcome of respondent *i* born in province *a* in year *y*. Three dependent variables are used separately. These include the respondent's adult height, the highest level of education claimed to have been attained, and monthly household consumption per capita at the time of the interview. As previously stated, we employ two distinct methods for identifying the respondents who were exposed to electoral violence during their early age, *Victim growing age*. The first proxy is a dummy variable that indicates whether the respondents self-reported being exposed to electoral violence directly (themselves or their households) or indirectly (their village) during their growing years. The second alternative proxy is a dummy variable that indicates whether the now-adult respondents were living in districts affected by electoral violence during 2007–08 during their growing age, from gestation up to 16 years old. Thus, the coefficient of interest, $\beta$, measures the difference in outcomes between respondents who were exposed to electoral violence while growing up and adult respondents who were not affected by electoral violence during their early life.

We also account for the characteristics of the respondents and their parents, denoted by vector *X*, to reduce potential biases. This vector includes the respondent's sex, whether they were bedridden for more than one month during their growing years, if they had malaria during their growing years, and whether they live in a rural or urban setting. Furthermore, we control for whether the respondents' parents are of the same ethnic group, their fathers' educational attainment, wealth index, and the number of cows owned during the respondents' growing years. We also include three fixed effects: respondent's birth cohort $\alpha_{iy}$, ethnicity $\varphi_{ie}$, and the province of birth $\mu_{ia}$. These fixed effects account for common time trends as well as unobserved ethnic and geographic heterogeneity, such as macroeconomic shocks, distinct ethnic group eating habits, average height, and local labor market conditions. The robust errors are clustered at the respondents' ethnicity and birth year levels.

*5.1 Coarsened matching*

Electoral violence does not occur at random. Despite the wide range of controls used, our ability to infer causal effects from our benchmark specifications is limited. Instead, we employ the coarsened matching estimator as our primary identification strategy. This estimator calculates the Average Treatment Effect on the Treated (ATT), in our case, experiencing electoral violence. This non-parametric estimator, proposed by Iacus et al. (2008), does not require to calculate propensity scores. Instead, it involves pruning observations ex-ante, so the remaining treatment and controls have a better balance of covariates in terms of marginal and joint distributions, ensuring that imbalances on one variable do not influence the maximum imbalance on other variables. Reducing imbalances in covariates minimizes model misspecification and potential biases resulting from unmeasured confounding factors. In turn, the estimator reduces within-group variation and improves treatment effect estimation precision. Although some observations with no close matches in variables may be excluded, matching ensures that all individuals within each pattern contribute to the analysis, thereby preserving a larger sample size than other methods such as propensity score matching.

In the matching procedure, we include variables that are likely to be significant confounding factors of electoral violence. These include whether the respondents were bedridden for more than a month during their childhood or adolescence, whether they live in a rural area, whether their parents are from the same ethnic group, and their father's educational attainment. After matching, the Average

Treatment Effect on the Treated (ATT) of exposure to electoral violence is estimated using OLS regression while controlling for any differences in covariates between the treatment and control groups. As suggested by Blackwell, Iacus, King, and Porro (2009), we also weigh this regression using the estimated coarsened matching weights to account for any remaining potential imbalance. We use robust errors clustered at the ethnicity and birth year levels of respondents.

## 6. Results

*6.1 OLS results*

Using the OLS regressions shown in Eq. (1), we examine all adult respondents who were exposed to electoral violence during their prenatal, childhood, or adolescence. This subsample consists of 840 respondents between the ages of 18 and 37 who were born during the highlighted cohorts listed in Table 1. We begin by analyzing the effect electoral violence had on respondents who were either directly (they or their households) or indirectly (their village) exposed to it. These respondents who were exposed to electoral violence during their early years are, on average, about three centimeters shorter than adults who were not exposed to electoral violence during their early years (Table 3, column 1).

These findings are consistent if we exclude from the analysis all respondents who grew up in districts that also experienced severe adverse agricultural production shocks due to extreme weather, rather than disruption caused by electoral violence (Table 3, column 2). We identify these districts using D'Alessandro et al. (2015)'s study, which has a comprehensive record of the districts that have suffered significant agricultural economic losses due to extreme weather conditions since the late 1980s.[3]

To test the robustness of our findings, we use our second alternative proxy to identify childhood or teen exposure to electoral violence. This time, we identify these respondents solely based on their date of birth and whether they resided in a district affected by the 2007–08 electoral violence. Using the data provided by (CIPEV, 2008), we identify the districts that experienced injuries or fatalities. At the time of their exposure to the electoral violence in 2007−08, these now-adult respondents were adolescents (12–13 years old).

This election in Kenya has been the most violent in the country's history, so it is important to examine these affected respondents. Furthermore, medical studies have shown that experiencing nutritional shocks during adolescence can still have detrimental effects on adult height. We find that respondents who grew up in districts where there was violence during the 2007–08 elections are about 3.9 cm shorter than people who grew up in districts unaffected by such violence (Table 3, column 3).

---

[3] We excluded from the analysis all people who lived in Narok in 1992, Garissa in 1997, Mombasa or Oi in 2002, Kisumu, Eldoret, Dagoretti, or Mombasa in 2007, or Mombasa in 2013.

**Table 3** Long-run effect of children's exposure to electoral violence on height, OLS results.

|  | (1) All sub-sample of respondents | (2) Removing districts that experienced extreme weather | (3) Identifying victims by district of residency during 2007-08 |
|---|---|---|---|
|  | Height | Height | Height |
| The respondent was victim of electoral violence back during growing age | -2.939** | -3.329** |  |
|  | (1.319) | (1.327) |  |
| The respondent grew in a district that experienced electoral violence during 2007-08 |  |  | -3.893* |
|  |  |  | (2.026) |
| The respondent is male | 7.654*** | 7.293*** | 8.093*** |
|  | (1.409) | (1.473) | (1.460) |
| The respondent was bedridden for more than one month when of growing age | -1.475 | -0.994 | -1.430 |
|  | (2.057) | (2.104) | (2.316) |
| The respondent got malaria when of growing age | 2.714* | 2.360 | 3.287* |
|  | (1.453) | (1.474) | (1.707) |
| The respondent got polio when of growing age | 6.016 | 5.820 | 6.643 |
|  | (4.312) | (4.444) | (5.541) |
| The respondent currently lives in a rural area | 1.038 | 1.233 | 0.942 |
|  | (1.187) | (1.205) | (1.317) |
| The respondent's parents belong to the same ethnic group | -1.259 | -1.171 | -0.655 |
|  | (1.218) | (1.203) | (1.409) |
| Number of cows the respondent's father owned back when the respondent was of growing age |  |  |  |
|  | 0.006 | -0.009 | 0.067 |
|  | (0.067) | (0.075) | (0.069) |
| The asset index of the respondent's father back when the respondent was of growing age | 8.472* | 8.811* | 8.043* |
|  | (4.583) | (4.600) | (4.719) |
| Respondent's ethnicity fixed effects | Yes | Yes | Yes |
| Respondent's province of birth fixed effects | Yes | Yes | Yes |
| Respondent's cohort of birth fixed effects | Yes | Yes | Yes |
| Educational attainment of respondents' father | Yes | Yes | Yes |
| Observations | 351 | 336 | 291 |
| R-squared | 0.288 | 0.291 | 0.316 |

In parentheses are the robust standard errors clustered at the ethnicity and birth year levels of the respondents. Significance levels: ***p < 0.01, **p < 0.05, *p < 0.1.

**Table 4** Long-run exposure to electoral violence on human capital and consumption, OLS results.

| | (1) | (2) | (3) | (4) | (5) | (6) |
|---|---|---|---|---|---|---|
| | All sub-sample | | Removing districts that experienced extreme weather | | Identifying victims by district of residency during 2007-08 | |
| | Educational attainment | Household consumption per capita | Educational attainment | Household consumption per capita | Educational attainment | Household consumption per capita |
| The respondent was victim of electoral violence back during growing age | 0.061 | -37.571 | 0.042 | -35.580 | | |
| | (0.111) | (44.611) | (0.112) | (46.463) | | |
| The respondent grew in a district that experienced electoral violence during 2007-08 | | | | | -0.194 | 10.792 |
| | | | | | (0.227) | (51.806) |
| The respondent is male | 0.156* | -9.936 | 0.119 | -10.748 | 0.074 | -27.841 |
| | (0.091) | (32.920) | (0.095) | (33.918) | (0.105) | (35.419) |
| The respondent was bedridden for more than one month when of growing age | -0.129 | -44.608* | -0.110 | -44.887* | -0.154 | -43.040* |
| | (0.175) | (22.687) | (0.179) | (23.574) | (0.189) | (25.186) |
| The respondent got malaria when of growing age | -0.048 | -52.661 | -0.097 | -53.003 | -0.009 | -42.279 |
| | (0.096) | (32.601) | (0.099) | (35.255) | (0.102) | (37.988) |
| The respondent got polio when of growing age | 0.272 | 59.732 | 0.276 | 55.812 | 0.482* | 58.897 |
| | (0.222) | (63.355) | (0.223) | (63.504) | (0.251) | (85.563) |
| The respondent currently lives in a rural area | -0.030 | 19.395 | -0.039 | 20.144 | 0.083 | -6.434 |
| | (0.098) | (34.400) | (0.102) | (35.597) | (0.109) | (35.425) |
| The respondent's parents belong to the same ethnic group | 0.052 | -30.724 | 0.041 | -35.696 | 0.065 | -18.206 |
| | (0.101) | (49.027) | (0.104) | (52.648) | (0.116) | (54.098) |
| Number of cows the respondent's father owned back when the respondent was of growing age | -0.001 | -0.083 | -0.001 | -0.089 | -0.005 | -0.123 |
| | (0.002) | (0.385) | (0.002) | (0.413) | (0.006) | (2.652) |
| The asset index of the respondent's father back when the respondent was of growing age | 1.913*** | 98.927 | 1.937*** | 113.797 | 1.750*** | 21.679 |
| | (0.302) | (167.183) | (0.309) | (181.331) | (0.346) | (209.932) |
| Respondent's ethnicity fixed effects | Yes | Yes | Yes | Yes | Yes | Yes |
| Respondent's province of birth fixed effects | Yes | Yes | Yes | Yes | Yes | Yes |
| Respondent's cohort of birth fixed effects | Yes | Yes | Yes | Yes | Yes | Yes |
| Educational attainment of respondents' father | Yes | Yes | Yes | Yes | Yes | Yes |
| Observations | 427 | 282 | 408 | 270 | 349 | 234 |
| R-squared | 0.360 | 0.158 | 0.346 | 0.158 | 0.381 | 0.164 |

In parentheses are the robust standard errors clustered at the ethnicity and birth year levels of the respondents. Significance levels: ***$p < 0.01$, **$p < 0.05$, *$p < 0.1$.

We continue to estimate the long-term impact of early life exposure to electoral violence on the respondents' contemporaneous educational attainment and their monthly household consumption per capita. First, we present the OLS estimates. There are no statistically significant differences between victims (those exposed to electoral violence during their growing years) and non-victims in terms of educational attainment or household consumption per capita (Table 4).[4] These results remain consistent, even after excluding respondents who grew up in districts that experienced extreme falls in agricultural production due to extreme weather (Table 4, columns 3–4). We also analyze the effects of electoral violence on the now-adult respondents, who were both of growing age and living in districts affected by electoral violence during 2007–08. These respondents have no statistically significant differences in educational attainment, or household consumption per capita compared to respondents growing up in districts unaffected by the violence (Table 4, columns 5–6).

---

[4] We also tried different educational attainment specifications, such as focusing solely on primary, secondary, or higher education. Again, we found no evidence that those affected by electoral violence had lower educational attainment than non-victims, so we will not present these findings here, but they are available upon request.

*6.2 Coarsened matching results*

To infer causal effects, we now concentrate on the coarsened matching estimator results. This method computes the Average Treatment Effect on the Treated (ATT) of being exposed to electoral violence by matching comparable respondents who were exposed to electoral violence during their growing years to those who were not. Table 5 shows the characteristics of respondents before and after they were matched. The bottom rows show the number of (matched) treated, controls, and the $L_1$ statistic, proposed by Iacus, King, and Porro (2008). This statistic measures the overall imbalance of all pretreatment covariates in the treatment and control groups, including full joint distributions and all interactions across all variables. This statistic ranges from 0 to 1, with 0 indicating no imbalances and 1 indicating complete separation of treatment and control groups. The first column of Table 5 shows the $L_1^j$ statistic which measures the $L_1$ for each variable separately, excluding interactions with other variables. The second column reports the difference in means, and the remaining columns show the difference in the quantiles of the distribution between the two groups for the 0th (min), 25th, etc., until the 100th (max percentile) for each variable. In our case, after matching the $L_1^j$ statistic is reduced to 0 for each variable, and the overall $L_1$ statistic is reduced from 0.42 to 0.25.

**Table 5** Adult respondents who were exposed or not to electoral violence during their growing age. Coarsened matching results.

| | Before matching | | | | | | | After matching | | | | | | |
|---|---|---|---|---|---|---|---|---|---|---|---|---|---|---|
| | L1 | mean | min | 25% | 50% | 75% | max | L1 | mean | min | 25% | 50% | 75% | max |
| The respondent was sick and bedridden for more than one month when of growing age | 0.05 | 0.05 | 0.00 | 0.00 | 0.00 | 0.00 | 0.00 | 0.00 | 0.00 | 0.00 | 0.00 | 0.00 | 0.00 | 0.00 |
| Educational attainment of the respondents' father | | | | | | | | | | | | | | |
| Some primary schooling or less | 0.11 | -0.11 | 0.00 | 0.00 | 0.00 | -1.00 | 0.00 | 0.00 | 0.00 | 0.00 | 0.00 | 0.00 | 0.00 | 0.00 |
| Primary school completed | 0.01 | -0.01 | 0.00 | 0.00 | 0.00 | 0.00 | 0.00 | 0.00 | 0.00 | 0.00 | 0.00 | 0.00 | 0.00 | 0.00 |
| Some secondary school to post-secondary qualifications | 0.01 | 0.01 | 0.00 | 0.00 | 0.00 | 0.00 | 0.00 | 0.00 | 0.00 | 0.00 | 0.00 | 0.00 | 0.00 | 0.00 |
| Some university or beyond | 0.02 | -0.02 | 0.00 | 0.00 | 0.00 | 0.00 | 0.00 | 0.00 | 0.00 | 0.00 | 0.00 | 0.00 | 0.00 | 0.00 |
| The parents of the respondent belong to the same ethnic group | 0.02 | -0.02 | 0.00 | 0.00 | 0.00 | 0.00 | 0.00 | 0.00 | 0.00 | 0.00 | 0.00 | 0.00 | 0.00 | 0.00 |
| The respondent currently lives in a rural area | 0.01 | 0.01 | 0.00 | 0.00 | 0.00 | 0.00 | 0.00 | 0.00 | 0.00 | 0.00 | 0.00 | 0.00 | 0.00 | 0.00 |
| The respondent is male | 0.02 | -0.02 | 0.00 | 0.00 | 0.00 | 0.00 | 0.00 | 0.00 | 0.00 | 0.00 | 0.00 | 0.00 | 0.00 | 0.00 |
| Respondent's ethnicity | | | | | | | | | | | | | | |
| Kikuyu | 0.09 | -0.09 | 0.00 | 0.00 | 0.00 | 0.00 | 0.00 | 0.00 | 0.00 | 0.00 | 0.00 | 0.00 | 0.00 | 0.00 |
| Luo | 0.05 | 0.05 | 0.00 | 0.00 | 0.00 | 0.00 | 0.00 | 0.00 | 0.00 | 0.00 | 0.00 | 0.00 | 0.00 | 0.00 |
| Other | 0.04 | 0.04 | 0.00 | 0.00 | 0.00 | 0.00 | 0.00 | 0.00 | 0.00 | 0.00 | 0.00 | 0.00 | 0.00 | 0.00 |
| Overall multivariate $L_1$ distance before matching | 0.42 | | | | | | | | | | | | | |
| Number of adult respondents in the control group before matching | 605 | | | | | | | | | | | | | |
| Number of adult respondents in the treatment group before matching | 235 | | | | | | | | | | | | | |
| Total number of observations, before matching | 840 | | | | | | | | | | | | | |
| Overall multivariate $L_1$ distance after matching | | | | | | | | 0.25 | | | | | | |
| Number of adult respondents in the control group after matching | | | | | | | | 422 | | | | | | |
| Number of adult respondents in the treated group after matching | | | | | | | | 212 | | | | | | |
| Total number of observations, after matching | | | | | | | | 634 | | | | | | |

$L_1$ represents the overall imbalance statistic proposed by Iacus, King, and Porro (2008).

**Table 6** ATT of respondents affected by electoral violence during their growing age. Coarsened matching results.

| | (1) Height | (2) Educational attainment | (3) Household consumption per capita |
|---|---|---|---|
| All matched sub-samples | | | |
| Panel A: Victims of electoral violence during growing age | | | |
| ATT of being exposed to electoral violence | -2.425* | 0.015 | -16.698 |
| | (1.352) | (0.124) | (22.603) |
| Respondent's ethnicity fixed effects | Yes | Yes | Yes |
| Respondent's province of birth fixed effects | Yes | Yes | Yes |
| Respondent's cohort of birth fixed effects | Yes | Yes | Yes |
| Other respondents' controls | Yes | Yes | Yes |
| Characteristics of the respondents' father | Yes | Yes | Yes |
| Observations | 471 | 609 | 405 |
| Removing districts that suffered extreme weather | | | |
| ATT of being exposed to electoral violence | -2.612* | 0.006 | -12.987 |
| | (1.430) | (0.127) | (24.445) |
| Respondent's ethnicity fixed effects | Yes | Yes | Yes |
| Respondent's province of birth fixed effects | Yes | Yes | Yes |
| Respondent's cohort of birth fixed effects | Yes | Yes | Yes |
| Other respondents' controls | Yes | Yes | Yes |
| Characteristics of the respondents' father | Yes | Yes | Yes |
| Observations | 449 | 583 | 387 |
| Panel B: Identifying victims of electoral violence by district of residency during 2007-08 | | | |
| ATT of being exposed to electoral violence | -4.966** | 0.020 | -25.660 |
| | (2.353) | (0.217) | (28.068) |
| Respondent's ethnicity fixed effects | Yes | Yes | Yes |
| Respondent's province of birth fixed effects | Yes | Yes | Yes |
| Respondent's cohort of birth fixed effects | Yes | Yes | Yes |
| Other respondents' controls | Yes | Yes | Yes |
| Characteristics of the respondents' father | Yes | Yes | Yes |
| Observations | 185 | 244 | 171 |

Respondents are matched based on their ethnicity, sex, whether they were bedridden for more than a month as children, rural residence, whether their parents are from the same ethnic group, and their parents' educational attainment. The regressions are weighted by the resulting coarsened weights. The regressions control for respondents' characteristics, including their sex, ethnicity, province of residence, cohort of birth, and whether they were affected by polio during childhood. The regressions also include the wealth index of the respondent's father, back when the respondent was of growing age. In parentheses are the robust standard errors clustered at the ethnicity and birth year levels of the respondents. Significance levels: ***p < 0.01, **p < 0.05, *p < 0.1.

We find that people who were exposed to electoral violence when they were young (from in utero to age 16) are almost 3 centimeters shorter than people who were not exposed to this kind of violence (Table 6). Nonetheless, we find no statistically significant impact on respondents' present educational attainment or monthly household consumption per capita. These results are robust if we remove respondents growing up in districts affected by extreme weather (Table 6, Panel A). In panel B, we estimate the ATT of adult respondents who grew up in districts where there was election violence in 2007–08. The results show that the heights of respondents who were exposed to election violence in 2007–08 are almost 5 cm shorter than the heights of similar respondents who grew up in other districts where they were not exposed to such violence (Table 6, Panel B).

We conducted additional tests to determine whether males and females have different long-run effects. For example, we included an interaction between the respondent's sex and having been exposed to electoral violence as a child or teen. These interactions suggest that there are no statistically significant differences between men and women; thus, are not included in Table 6.

*6.3 Placebo tests: exposed to electoral violence during adulthood*

We present next two placebo tests which help us rule out the possibility that our findings are the result of chance, pre-trends, or important unobserved confounding factors. The coarsened estimator is used to estimate both placebo tests. The first placebo test estimates the ATT on respondents who were exposed to electoral violence after they had stopped growing, between the ages of 18 and 60. Thus, any food shortage or diet change experienced at the time should not affect their height or other related factors when comparing respondents who went through electoral violence as adults with similar respondents who did not experience it. To estimate this ATT, we compare the outcomes of non-growing-age victims of electoral violence to non-victims of electoral violence who were no longer growing at the time of the violence. As expected, we find no impact on the height, current educational attainment, or household consumption per capita of these victims of electoral violence who had stopped growing at the time of the event (Table 7, Panel A). If we exclude all respondents from this placebo test who resided in areas that experienced extreme weather, we consistently obtain the same results (Panel A).

As a second placebo test, we compare respondents who were living in districts that experienced electoral violence in 2007–08 to residents of districts that did not experience electoral violence during their non-growing years. As expected, exposure to electoral violence had no long-term effects on these respondents' adult height, educational attainment, or monthly household consumption per capita (Table 7, Panel B).

**Table 7** Placebo tests: ATT of respondents who, during their non-growing years, were affected by electoral violence. Coarsened matching results.

| | (1) Height | (2) Educational attainment | (3) Household consumption per capita |
|---|---|---|---|
| All matched sub-samples | | | |
| *Placebo A: Victim of electoral violence during non-growing age* | | | |
| ATT of being exposed to electoral violence | -1.634 | -0.115 | -32.688 |
| | (2.912) | (0.202) | (22.019) |
| Respondent's ethnicity fixed effects | Yes | Yes | Yes |
| Respondent's province of birth fixed effects | Yes | Yes | Yes |
| Respondent's cohort of birth fixed effects | Yes | Yes | Yes |
| Other respondents' controls | Yes | Yes | Yes |
| Characteristics of the respondents' father | Yes | Yes | Yes |
| Observations | 155 | 192 | 131 |
| *Removing districts that suffered extreme weather* | | | |
| ATT of being exposed to electoral violence | -1.634 | -0.115 | -32.688 |
| | (2.912) | (0.202) | (22.019) |
| Respondent's ethnicity fixed effects | Yes | Yes | Yes |
| Respondent's province of birth fixed effects | Yes | Yes | Yes |
| Respondent's cohort of birth fixed effects | Yes | Yes | Yes |
| Other respondents' controls | Yes | Yes | Yes |
| Characteristics of the respondents' father | Yes | Yes | Yes |
| Observations | 155 | 167 | 114 |
| *Placebo B: Living in districts that experienced electoral violence in 2007-08 during non-growing age* | | | |
| ATT of being exposed to electoral violence | -5.261 | -0.222 | 24.194 |
| | (4.265) | (0.313) | (29.506) |
| Respondent's ethnicity fixed effects | Yes | Yes | Yes |
| Respondent's province of birth fixed effects | Yes | Yes | Yes |
| Respondent's cohort of birth fixed effects | Yes | Yes | Yes |
| Other respondents' controls | Yes | Yes | Yes |
| Characteristics of the respondents' father | Yes | Yes | Yes |
| Observations | 142 | 177 | 124 |

Respondents are matched based on their ethnicity, sex, whether they were bedridden for more than a month as children, rural residence, whether their parents are from the same ethnic group, and their parents' educational attainment. The regressions are weighted by the resulting coarsened weights. The regressions control for respondents' characteristics, including their sex, ethnicity, province of residence, cohort of birth, and whether they were affected by polio during childhood. The regressions also include the wealth index of the respondent's father, back when the respondent was of growing age. In parentheses are the robust standard errors clustered at the ethnicity and birth year levels of the respondents. Significance levels: ***p < 0.01, **p < 0.05, *p < 0.1.

## 7. Second-generation impacts

We now turn our attention to the paper's second research question. That is, whether exposure to electoral violence as a child or adolescent affects the victims' descendants. To test for this second-generation effect, we examine whether electoral violence affected the children of respondents who experienced electoral violence during their growing years. To do so, we compare the height-for-age z-scores of children of electoral violence victims to those of children of non-victims of electoral violence. To begin, we use the OLS regression shown in Eq. (2) to estimate the second-generation effect.

*child outcome$_{cay}$* = $\beta_1$ *Child of victim$_{i,y}$* + $\beta_2$ *Boy\*Child of victim$_{i,y}$* + $\eta C_c$ + $\alpha_{cy}$ + $\varphi_{ce}$ + $\mu_{ca}$ + $\varepsilon_c$  (2)

The dependent variable, *child outcome$_{cay}$*, represents the analyzed impact of the respondent's child *c* born in the province *a* in year *y*. We use two dependent variables separately: educational attainment and height-for-age of the children, measured using the so-called z-scores. These z-scores have been standardized to be gender and age-neutral. That is, height can be compared between girls and boys while accounting for sex and age differences in the population. Thus, z-scores enable the evaluation of children's growth status by putting together children regardless of their sex and age, ranging from newborns to adults. *Child of victim$_i$* indicates whether the child's parent *i* (mother or father) was affected by electoral violence during her/his growing years in the province *a*. To test whether there are any gender differences in the second-generation effects, in a second specification, we add the interaction on whether the child is a boy and is a child of a respondent who was exposed to electoral violence while growing up (*Boy\*Child of victim*). The respective coefficient $\beta_2$ quantifies the impact differential between sons and daughters of respondents who were exposed to electoral violence during prenatal, childhood, or adolescence, relative to the height difference between sons and daughters of respondents who were not exposed to the violence.

To reduce the possibility of a self-selection bias, we control for vector *C*, which includes the sex of the children, whether they have been bedridden for more than a month, and whether they reside in a rural area. In this vector, we also include the characteristics of the child's parents, such as the mother's or father's height, educational attainment, employment status, and birth cohort (depending on who was interviewed). In addition, we control for whether the parents of the children belong to the same ethnic group, their household wealth index, and the number of cows owned by their household at the time of the interview. We control for the height of the parents because stunted parents may be more likely to have children with low height-for-age z-scores. All the other terms on the right-hand side refer to the child's birth cohort $\alpha_{cy}$, ethnicity $\varphi_{ce}$, and the province of residence $\mu_{ca}$. These fixed effects capture common time trends and unobserved heterogeneity at the ethnicity and area levels, including macroeconomic shocks, the eating habits of ethnic groups, their average height, and local labor market conditions. Because the majority of families have multiple children, robust errors are clustered at the children's ethnicity, birth year, and household levels.

*7.1 OLS results*

To determine whether second-generation effects are long-lasting, we compare the height and educational attainment of the children of adult respondents who were or were not exposed to electoral violence during their growing years. We focus on our first proxy measure on whether adult respondents were exposed directly or indirectly to electoral violence during their growing years.

In column 1 of Table 8 is shown the first model specification. There is no difference in height-for-age z-scores between the children (girls and boys) of respondents who experienced electoral violence during their growing years and the children of respondents not exposed to such violence. In column 2 of Table 8, the second model specification is shown, which includes the interaction *Children of victim electoral violence\*Boys.* This interaction suggests that boys are more stunted than girls among the children of the respondents who were affected by electoral violence during growing age, relative to the height difference between sons and daughters of respondents who were not affected by the violence. To clarify this finding further, we restrict the sample to male children in column 3. This third specification reveals that the sons of adult respondents who experienced electoral violence during their growing years also have lower height-for-age z scores (-3.40 units) than the sons of respondents who did not experience such violence. If the sample is restricted to girls only, in column 4, we find no such effect for daughters.

We also compare the educational attainment of children whose parents were victims of electoral violence to the educational attainment of children whose parents were not victims of electoral violence. We concentrate on school-aged children (ages 6 to 18). This sample restriction is consistent with the relevant literature, which suggests that excluding children younger than the school-entry age helps to reduce any potential selectivity among children remaining at home (Akresh et al., 2021). There is no evidence that the educational attainment of the children of the victims of electoral is lower than that of the children of non-victims (column 5). We find no difference in educational attainment between the sons and daughters of adult respondents who were exposed to electoral violence during their growing years (column 6). If we restrict the analysis to boys only, or girls only, we find no difference in educational attainment between the children of victims and non-victims of electoral violence (columns 7 and 8).

**Table 8** Height and educational attainment of the children of those victims of electoral violence who were affected during prenatal, childhood, or adolescence. OLS regression results.

| | (1) | (2) | (3) | (4) | (5) | (6) | (7) | (8) |
|---|---|---|---|---|---|---|---|---|
| | Height-for-age z score of children aged 18 and younger | | | | Educational attainment of students aged 6-18 | | | |
| | All sub-sample | Boys only | Girls only | | All sub-sample | Boys only | Girls | |
| Child of an adult respondent who was victim of electoral violence during prenatal, childhood, or adolescence | -0.531 | 0.323 | -3.407** | 0.851 | 0.265 | 0.042 | 1.098 | 0.632 |
| | (0.739) | (0.732) | (1.512) | (1.031) | (0.604) | (0.718) | (1.450) | (1.152) |
| Child of a victim of electoral violence back when of growing age X Male child | | -2.867*** | | | | 0.562 | | |
| | | (0.737) | | | | (0.950) | | |
| Male child | 0.044 | 0.655 | | | -0.561 | -0.653 | | |
| | (0.371) | (0.409) | | | (0.375) | (0.432) | | |
| Child has been bedridden for more than one month | -0.027 | 0.403 | 0.087 | -0.291 | 0.911 | 0.850 | 0.246 | 1.073 |
| | (0.696) | (0.680) | (2.144) | (1.210) | (1.132) | (1.136) | (1.444) | (2.532) |
| Number of cows owned by the child's household | 0.152 | 0.191 | -0.456 | 0.334* | -0.234* | -0.236* | -0.146 | -0.039 |
| | (0.170) | (0.163) | (0.531) | (0.196) | (0.121) | (0.123) | (0.264) | (0.206) |
| Child's parent is currently employed | -0.664 | -0.558 | -1.747* | 0.343 | 0.092 | 0.058 | 0.340 | 0.209 |
| | (0.507) | (0.467) | (0.975) | (0.636) | (0.450) | (0.459) | (1.012) | (0.931) |
| Household asset index of the child's household | 2.338 | 1.435 | 5.296 | 0.653 | 2.112 | 2.287 | 3.188 | 1.764 |
| | (2.057) | (1.933) | (5.546) | (2.652) | (1.600) | (1.656) | (2.968) | (3.245) |
| Height of the child's parent | 0.019 | 0.020 | 0.039 | 0.015 | 0.041 | 0.041 | 0.045 | 0.064 |
| | (0.028) | (0.026) | (0.074) | (0.051) | (0.027) | (0.027) | (0.060) | (0.057) |
| Currently residing in a rural area | 0.022 | -0.051 | -0.988 | 0.060 | -1.117* | -1.113* | -0.906 | -2.460* |
| | (0.472) | (0.433) | (0.890) | (0.755) | (0.613) | (0.618) | (1.280) | (1.244) |
| Child's ethnicity fixed effects | Yes | Yes | Yes | Yes | Yes | Yes | Yes | Yes |
| Child's province of residence fixed effects | Yes | Yes | Yes | Yes | Yes | Yes | Yes | Yes |
| Child's cohort of birth | Yes | Yes | Yes | Yes | Yes | Yes | Yes | Yes |
| Child parent's cohort of birth | Yes | Yes | Yes | Yes | Yes | Yes | Yes | Yes |
| Educational attainment of child's father | Yes | Yes | Yes | Yes | Yes | Yes | Yes | Yes |
| Observations | 111 | 111 | 50 | 61 | 132 | 132 | 66 | 66 |
| R-squared | 0.378 | 0.456 | 0.680 | 0.560 | 0.655 | 0.656 | 0.707 | 0.781 |

In parentheses are the robust standard errors clustered at the child's ethnicity, birth year, and household levels. Significance levels: ***p < 0.01, **p < 0.05, *p < 0.1.

*7.2 Coarsened matching results*

To infer causal effects, we then use coarsened matching. As before, we compare children of adult respondents whose early years were impacted by electoral violence to children of adult respondents whose early years were not impacted by such violence. Children are matched based on their ethnicity, sex, birth cohort, and parents' wealth index. Table 9 shows the balancing statistics before and after the match. Following matching, we estimate the ATT using OLS regressions weighted by the coarsened weights. The regressions include controls for children's characteristics, including their ethnicity, sex, province of residence, birth cohort, and whether they have been bedridden for more than one month, as well as their household's wealth index. Other controls include whether the child's parent interviewed is currently employed, her or his level of education, height, and birth cohort. The robust standard errors are clustered at the child's ethnicity, birth year, and household levels.

**Table 9** Children whose parents were or were not exposed to electoral violence during their prenatal, childhood, or adolescence. Coarsened matching results.

| | Before matching | | | | | | | After matching | | | | | | |
|---|---|---|---|---|---|---|---|---|---|---|---|---|---|---|
| | L1 | mean | min | 25% | 50% | 75% | max | L1 | mean | min | 25% | 50% | 75% | max |
| Child's cohort | 0.04 | 0.08 | 0.00 | 0.00 | 1.00 | 0.00 | 0.00 | 0.02 | 0.03 | 0.00 | 0.00 | 0.00 | 0.00 | 0.00 |
| Child's sex | 0.10 | -0.10 | 0.00 | 0.00 | -1.00 | 0.00 | 0.00 | 0.00 | 0.00 | 0.00 | 0.00 | 0.00 | 0.00 | 0.00 |
| Household asset index of the child's household | 0.16 | -0.02 | 0.00 | 0.00 | 0.00 | 0.00 | -0.13 | 0.06 | -0.01 | 0.00 | 0.00 | 0.00 | 0.00 | 0.00 |
| Ethnicity | | | | | | | | | | | | | | |
|   Kikuyu | 0.04 | -0.04 | 0.00 | 0.00 | 0.00 | 0.00 | 0.00 | 0.00 | 0.00 | 0.00 | 0.00 | 0.00 | 0.00 | 0.00 |
|   Luo | 0.03 | -0.03 | 0.00 | 0.00 | 0.00 | 0.00 | 0.00 | 0.00 | 0.00 | 0.00 | 0.00 | 0.00 | 0.00 | 0.00 |
|   Other | 0.07 | 0.07 | 0.00 | 1.00 | 0.00 | 0.00 | 0.00 | 0.00 | 0.00 | 0.00 | 0.00 | 0.00 | 0.00 | 0.00 |
| Overall multivariate $L_1$ distance before matching | 0.468 | | | | | | | | | | | | | |
| Number of children in the control group before matching | 318 | | | | | | | | | | | | | |
| Number of children in the treatment group before matching | 98 | | | | | | | | | | | | | |
| Total observations before matching | 416 | | | | | | | | | | | | | |
| Overall multivariate $L_1$ distance after matching | | | | | | | | 0.334 | | | | | | |
| Number of children in the control group after matching | | | | | | | | 269 | | | | | | |
| Number of treated children in the treatment group after matching | | | | | | | | 97 | | | | | | |
| Total observations after matching | | | | | | | | 366 | | | | | | |

$L_1$ represents the overall imbalance statistic proposed by Iacus, King, and Porro (2008).

After matching the children of the respondents, the ATT is estimated using an OLS regression that takes into account differences between the treatment and control groups. We use the weights generated by the coarsened matching method to this regression to mitigate any potential remaining imbalance. This regression also controls for children's characteristics including their sex, ethnicity, province of residence, cohort of birth, whether they have been bedridden for more than one month, and household wealth index. Other controls include whether the parent of the child interviewed is currently employed, educational attainment, height, and cohort of birth. We cluster the robust standard errors at the child's ethnicity, birth year, and household levels.

Table 10, column 1, shows the ATT for children aged 18 or younger. Panel A shows that there is no difference in height-for-age z-scores between the children (boys and girls) of respondents who were exposed to electoral violence when they were growing up and the children of respondents not exposed to such violence. But in the second model specification, Panel B, we find that there is an effect on boys. This second specification includes the interaction *Children of victim electoral violence*Boys.* As before, we find that the sons of respondents whose growing years were impacted by electoral violence have statistically significantly lower height-for-age z scores than daughters of such parents, relative to the height difference between sons and daughters of parents not affected by electoral violence. To clarify this result further, we restrict the sample in Panel C to boys only. This third specification reveals that the sons of adult respondents who were exposed to electoral violence during their growing years are shorter in height than the sons of respondents who were not exposed to such violence. Panel D shows the results if we restrict the attention to girls only. We find no statistically

significant differences in height-for-age z scores or educational attainment between the daughters of those who experienced electoral violence during their growing years and those who did not.

As in major and protracted civil wars, our findings indicate that stunting is transmitted from one generation to the next due to the significant shocks that households experience during electoral violence. In Kenya, however, the small-scale electoral violence has only stunted the growth of boys in the second generation. Why has this stunting effect been passed down to male descendants only? There may be a number of different mechanisms at play, but one of them may undoubtedly be how nutrition shocks affect boys and girls. There is evidence that boys in Kenya are more likely to be stunted (30%) than girls (22%), as measured by z-scores below two standard deviations (Kenya National Bureau of Statistics 2015, p.187). Multiple countries in the region have a higher incidence of male stunting. Even without civil conflict, two large meta-analyses from Sub-Saharan Africa indicate that boys are more likely than girls to be stunted (Keino et al., 2014; Kenyan Ministry of Health, 2011; Wamani et al., 2007). To date, the medical or socioeconomic cause of the observed gender disparity in stunting in the region remains unknown (Vonaesch et al., 2017). A possible explanation for these disparities in stunting is the difference in caloric requirements and associated costs between boys and girls. Boys and girls have comparable caloric needs during childhood. Boys' caloric requirements increase more rapidly than girls' during adolescence, reaching the same level as adults or even higher if they are extremely active (Public Health England, 2018). Due to these disparities, boys may be more susceptible to consumption shocks in the home than girls.

**Table 10** ATT of children whose parents were affected by electoral violence during their prenatal, childhood, or adolescence. Coarsened matching results.

| | (1) Height-for-age z score of children aged 18 or younger | (2) Height-for-age z score of children aged 14 or younger | (3) Educational attainment of school-aged children 6-18 |
|---|---|---|---|
| **Panel A: Matched children of non-victims and victims who were exposed to electoral violence during their growing age** | | | |
| ATT of being a child of a respondent who was victim of electoral violence during prenatal, childhood, or adolescence | -0.508 | -0.535 | -0.118 |
| | (0.698) | (0.696) | (0.385) |
| Child's ethnicity fixed effects | Yes | Yes | Yes |
| Child's province of residence fixed effects | Yes | Yes | Yes |
| Child's cohort of birth | Yes | Yes | Yes |
| Other child's controls | Yes | Yes | Yes |
| Parent's cohort of birth | Yes | Yes | Yes |
| Parent's controls | Yes | Yes | Yes |
| Observations | 116 | 114 | 235 |
| **Panel B: Amongst children matched testing for differential effects between boys and girls** | | | |
| ATT of being a son of a respondent who was victim of electoral violence during prenatal, childhood or adolescence | -3.155*** | -3.281*** | 0.724 |
| | (0.773) | (0.768) | (0.602) |
| Child's ethnicity fixed effects | Yes | Yes | Yes |
| Child's province of residence fixed effects | Yes | Yes | Yes |
| Child's cohort of birth | Yes | Yes | Yes |
| Other child's controls | Yes | Yes | Yes |
| Parent's cohort of birth | Yes | Yes | Yes |
| Parent's controls | Yes | Yes | Yes |
| Observations | 116 | 114 | 235 |
| **Panel C: Comparing boys only. Sons of victims and sons of non-victims of electoral violence who were exposed to the violence during their growing years** | | | |
| ATT of being the son of a respondent who was victim of electoral violence during prenatal, childhood or adolescence (relative to sons of non-victims) | -2.938** | -2.943** | 0.327 |
| | (1.442) | (1.430) | (0.585) |
| Child's ethnicity fixed effects | Yes | Yes | Yes |
| Child's province of residence fixed effects | Yes | Yes | Yes |
| Child's cohort of birth | Yes | Yes | Yes |
| Other child's controls | Yes | Yes | Yes |
| Parent's cohort of birth | Yes | Yes | Yes |
| Parent's controls | Yes | Yes | Yes |
| Observations | 53 | 52 | 119 |
| **Panel D: Comparing girls only. Daughters of victims and daughters of non-victims of electoral violence who were exposed to the violence during their growing years** | | | |
| ATT of being the daughter of a respondent who was victim of electoral violence during prenatal, childhood or adolescence (relative to daughters of non-victims) | 0.999 | 1.000 | -0.287 |
| | (0.912) | (0.912) | (0.601) |
| Child's ethnicity fixed effects | Yes | Yes | Yes |
| Child's province of residence fixed effects | Yes | Yes | Yes |
| Child's cohort of birth | Yes | Yes | Yes |
| Other child's controls | Yes | Yes | Yes |
| Parent's cohort of birth | Yes | Yes | Yes |
| Parent's controls | Yes | Yes | Yes |
| Observations | 61 | 61 | 112 |

Children are matched based on their ethnicity, sex, and the wealth index of their parents. The regressions control for children's characteristics including their sex, ethnicity, province of residence, cohort of birth, whether they have been bedridden for more than one month, and their household wealth index. Other controls include whether the parent of the child interviewed was currently employed, the parent's educational attainment, height, and cohort of birth. In parentheses are the robust standard errors clustered at the child's ethnicity, birth year, and household levels. Significance levels: ***$p < 0.01$, **$p < 0.05$, *$p < 0.1$.

As a robustness check, we examine the possibility of age-dependent heterogeneity in the effects of violence on the second generation. We examine children from birth to 14 years of age because there are no substantial differences between boys' and girls' caloric needs, nor associated gender-based cost differences for households. We find that sons of adult respondents whose early years were affected by electoral violence have lower z-scores than their daughters, relative to the height difference between sons and daughters of parents not affected by the violence (Table 10, column 2, Panel B). In Table 10, column 2, Panel C, only boys are included in the sample. Again, we find that the sons of adult respondents who were exposed to electoral violence during their formative years are shorter than the sons of adult respondents who were not exposed to such violence.

Another explanation for why boys are more stunted than girls is that households adjust their consumption patterns to protect daughters at the expense of sons. In the Sub-Saharan region, there is no strong evidence of preferential treatment for girls or sons (Friedman and Schady, 2009). With the rich retrospective data available for this study, it is impossible to test directly for such gender differences in household expenditures. However, one way to determine this is to compare the educational attainment of children whose parents were exposed to electoral violence during their childhood or adolescence. If households adapt their consumption to favor daughters over sons, we may see a difference in educational attainment. Again, we find no impact on the educational attainment of the children of respondents affected by electoral violence during their growing years (Table 10, column 3, Panel A). There are no educational attainment differences between boys and girls (Panel B). When only boys are considered and the sons of victims and non-victims of electoral violence are compared, there is no difference in educational attainment (Panel C). We also find no difference in the level of education between the daughters of people who were victims of electoral violence and those who were not (Panel D). The results of the second generation are robust to remove the districts that experienced extreme weather. The results are also robust if we focus on the children of respondents who lived in districts affected by the electoral violence in 2007–08. We do not present either of these results because the matched sample is reduced considerably, and might be insufficient to draw sound inferences.

The children who have been analyzed so far were born to parents who grew up during times of electoral violence. Some of these children are old enough, though, to have been indirectly affected by food price increases due to electoral violence. Next, we examine exclusively children born to parents who experienced violence during their childhood or adolescence, but who were born after the 2007–08 electoral violence. These children have not been subjected to significant food shocks as a result of violence and are thus a good way to test the effects of second-generation impacts.

**Table 11** ATT of children born after the 2007−08 electoral violence whose parents were affected by electoral violence during their prenatal, childhood, or adolescence. Coarsened matching results.

|  | (1) Height-for-age z score of children aged 6 or younger |
|---|---|
| Panel A: Matched children of non-victims and victims electoral violence during their growing age | |
| ATT of being a child of a respondent who was victim of electoral violence during prenatal, childhood, or adolescence | -1.475 |
|  | (1.256) |
| Child's ethnicity fixed effects | Yes |
| Child's province of residence fixed effects | Yes |
| Child's cohort of birth | Yes |
| Other child's controls | Yes |
| Parent's cohort of birth | Yes |
| Parent's controls | Yes |
| Observations | 59 |
| Panel B: Amongst children matched testing for differential effects between boys and girls | |
| ATT of being a son of a respondent who was victim of electoral violence during prenatal, childhood or adolescence | |
|  | -4.448*** |
|  | (1.317) |
| Child's ethnicity fixed effects | Yes |
| Child's province of residence fixed effects | Yes |
| Child's cohort of birth | Yes |
| Other child's controls | Yes |
| Parent's cohort of birth | Yes |
| Parent's controls | Yes |
| Observations | 59 |

Children are matched based on their ethnicity, sex, and the wealth index of their parents. The regressions control for children's characteristics including their sex, ethnicity, province of residence, cohort of birth, whether they have been bedridden for more than a month, and their household wealth index. Other controls include whether the parent of the child interviewed was currently employed, the parent's educational attainment, height, and cohort of birth. In parentheses are the robust standard errors clustered at the child's ethnicity, birth year, and household levels. Significance levels: ***$p < 0.01$, **$p < 0.05$, *$p < 0.1$.

The ATT for children aged six and younger is shown in Table 11. Panel A demonstrates that there is no difference in height-for-age z-scores between respondents' children (boys and girls) who were exposed to electoral violence when they were growing up and those who were not. Panel B of the second model specification, on the other hand, reveals that there is an effect on boys. This second specification contains the interaction *Children of victim electoral violence*Boys.* That is, sons of respondents whose growing years were affected by electoral violence have statistically lower height-for-age z scores than their daughters, relative to the height difference between sons and daughters of parents not affected by the violence. We do not examine the sample further for boys or girls because the sample size is significantly reduced. Nonetheless, the Table 11 results are reassuring because they suggest the existence of second-generation effects.

*7.2 Placebo tests among children of victims of electoral violence*

So far, we have shown that adult respondents who were affected by electoral violence when they were young do not have lower monthly household consumption per capita over time. But our way of measuring consumption might not be complete enough, and these families might still be poorer, which is why perhaps their children are shorter. If it is true that being a victim of violence lowers long-term consumption, which in turn lowers the height of children, then the children of people who were affected by electoral violence as adults should have the same effects. As our final robustness test, Table A.4 shows that the children of respondents who experienced electoral violence during adulthood have no lower height-for-age or lower educational attainment when compared to similar children of parents who did not experience electoral violence. We used the same coarsened technique and controls as before to match these children.

In sum, our findings indicate that exposure to electoral violence during prenatal, childhood, and adolescence is associated with a significant nutritional shock manifested by decreased height. In addition, this health shock is transmitted to the children of these victims of violence, particularly boys. Our first- and second-generation findings also indicate that electoral violence in Kenya has had no effect on the educational attainment of adults exposed to the violence during their growing years, nor on the educational attainment of their children. There are several potential reasons for these findings. In contrast to a number of other studies, our paper focuses on a small-scale conflict with a brief duration, which may help to mitigate its potential impact on educational attainment. Furthermore, Kenyan elections in 1992, 1997, 2002, and 2007 occurred during the December school holidays.[5] The timing of these elections could have prevented a substantial disruption in the provision of education in terms of opening hours, and potential teacher and student absenteeism caused by the violence.[6] Despite the number of households being displaced to camps for months (if not longer), most of these families could not return to their former homes and were forced to settle elsewhere, where they might have found alternative educational opportunities. Even though there are no primary school fees in Kenya since 2003, the country ranks ninth in the world for the number of children of primary school age who are not enrolled. This is due in part to a lack of schools and the persistence of high indirect school costs, which may affect many households, not just those exposed to electoral violence (UNESCO, 2012).

---

[5] Only the election of 2013 was conducted in March. This recent election did not affect our analyzed adult respondents during their growing age.

[6] Primary and secondary school academic years begin in January and end in November, with three months off in April, August, and December.

## 8. Conclusion

This paper examined the long-term health and human capital effects of prenatal, childhood, and adolescent exposure to electoral violence. It also examined whether these effects are transmitted to the offspring of these victims of violence. We conducted a nationally representative survey of adult respondents and their children in Kenya to answer these questions. Our case study has not experienced a civil war or a military coup. However, every single general election in the country has been marred by electoral violence since the reintroduction of multiparty elections in the early 1990s. This type of political violence affects numerous developing democracies, making our findings relevant to a wide range of similar settings.

In the first part of this paper, we examined the long-term consequences of being exposed to electoral violence during early life. We define this period from conception to the age of sixteen. We identified who had been affected by electoral violence by exploiting the temporal and spatial variation of electoral violence in Kenya between 1992 and 2013 in conjunction with a rich nationally representative. We also used official data to identify which districts were affected by the largest episode of electoral violence in the country in 2007–08. Using the coarsened matching estimator, we matched these now-adult respondents with comparable adults who were not exposed to such violence during their growing years. We matched respondents based on factors such as ethnicity, socioeconomic status, and parental characteristics. We found significant differences in adult height between those exposed to early-life electoral violence and those who were not. The now-adult respondents exposed to electoral violence during their childhood or adolescence are nearly three centimeters shorter than comparable adults who were not exposed to it.

We also examined the long-term effects of exposure to electoral violence on educational attainment. Impacts on human capital are another aspect that has been extensively examined in studies analyzing civil war and large-scale violence (Akresh et al., 2021; Alderman et al., 2006). In contrast to these earlier studies, we found no long-term effect of early-life exposure to electoral violence on educational attainment. The contrasting results between height and education may be explained by when electoral violence impacts the nation. Typically, elections are held in Kenya during the December holiday break from school. Consequently, significant economic and supply chain disruptions have a diminished impact on the educational cycle. That is, in contrast to other large-scale armed conflicts, our case study may not be susceptible to long-term disruptions in student and teacher attendance, thereby mitigating its impact on education. Our analysis lacks test scores to evaluate other educational effects of electoral violence. As an alternative test, we analyzed the long-term effect on household consumption per capita as a proxy for human capital accumulation during childhood and adolescence. We found no differences between those who were exposed to electoral violence in childhood or adolescence and those who were not.

In the second part of the paper, we investigated the effects on the second generation. We examined the height-for-age and educational attainment of children whose parents were impacted by electoral violence during their prenatal, childhood, or adolescence. We found an impact on the height-for-age of the descendants of electoral violence victims, as in other studies examining the effects of extensive and protracted civil wars (Akresh et al., 2021). Furthermore, we found that the sons of victims of electoral violence who were exposed to violence as children or teens are more stunted. We found no differences in educational attainment between children exposed to electoral violence early in life and those who were not. That is, the long-run effects of electoral violence in the first and second generations are only found on height.

These findings call for future research examining the long-term effects of violence beyond the immediate and important casualties and political outcomes. It is crucial to design more effective humanitarian responses to conflicts (Carroll et al., 2017; Onyango et al., 2019). Given the cyclical nature of electoral violence, emergency nutrition programs are crucial in regions with a well-known high risk of food shortages due to violence. Long-term interventions should address the observed disparities in the risk of stunting between boys and girls.

# Appendix

**Table A.1** Characteristics of the first-generation sample.

|  | Mean | St. Dev. |
|---|---|---|
| Female | 0.48 | 0.5 |
| Age 18-26 | 0.31 | 0.46 |
| Height for female respondents | 161.15 | 11.75 |
| Height for male respondents | 167.61 | 13.28 |
| Secondary or more for female respondents | 0.73 | 0.44 |
| Secondary or more for male respondents | 0.8 | 0.4 |
| Victim of electoral violence during growing age | 0.19 | 0.4 |
| Assent index | 0.41 | 0.17 |
| Monthly household consumption per capita (USD) | 72.82 | 234.96 |
| Employed | 0.5 | 0.5 |
| Married | 0.5 | 0.5 |
| Has children | 0.45 | 0.5 |
| Number of children | 0.74 | 1.13 |
| Lives in a rural area | 0.58 | 0.49 |
| Ethnicity | Freq. | |
|    Kikuyu | 16.94 | |
|    Luo | 15.03 | |
|    Luhya | 14.70 | |
|    Kamba | 10.22 | |
|    Meru | 7.89 | |
|    Kisii | 7.48 | |
|    Kalenjin | 10.13 | |
|    Maasai | 1.66 | |
|    Mijikenda | 6.06 | |
|    Taita | 2.66 | |
|    Somali | 3.41 | |
|    Pokot | 0.25 | |
|    Turkana | 1.58 | |
|    Teso | 0.25 | |
|    Embu | 1.33 | |
|    Other | 0.42 | |
| Province | | |
|    Nairobi | 7.19 | |
|    Central | 8.60 | |
|    Eastern | 16.20 | |
|    Rift Valley | 23.97 | |
|    Nyanza | 15.12 | |
|    Western | 12.73 | |
|    North Eastern | 3.22 | |
|    Coast | 12.98 | |
| When of growing age | Mean | St. Dev. |
| Respondent was bedridden for more than one month | 0.10 | 0.31 |
| Respondent got malaria | 0.56 | 0.50 |
| Respondent got polio | 0.03 | 0.17 |
| The characteristics of the respondent's father (or mother's characteristics if respondent never met his father) | Mean | St. Dev. |
| The household asset index of the respondent's father, back when the father was of the respondent's age | 0.39 | 0.17 |
| Number of cows owned by the respondent's father, back when the father was of the respondents' age | 6.16 | 14.03 |
| The respondent's father has secondary or more, back when the father was of the respondent's age | 0.45 | 0.50 |
| Observations | 1,210 | |

**Table A.2** Descriptive statistics of victims and non-victims of electoral violence (first-generation).

| | Experienced electoral violence during prenatal, childhood, or adolescence | | | Did not experience electoral violence during prenatal, childhood, or adolescence | | |
|---|---|---|---|---|---|---|
| | Obs. | Mean | St. Dev. | Obs. | Mean | St. Dev. |
| Current condition as adult | | | | | | |
| Height for female respondents | 77 | 159.97 | 9.43 | 228 | 161.96 | 11.54 |
| Height for male respondents | 88 | 167.08 | 11.95 | 239 | 168.91 | 12.07 |
| Secondary or more for female respondents | 107 | 0.79 | 0.41 | 292 | 0.79 | 0.41 |
| Secondary or more for male respondents | 121 | 0.86 | 0.35 | 302 | 0.80 | 0.40 |
| Assent index | 235 | 0.41 | 0.17 | 605 | 0.40 | 0.18 |
| Monthly household consumption per capita (USD) | 152 | 66.21 | 171.37 | 391 | 77.68 | 223.49 |
| Employed | 235 | 0.47 | 0.5 | 605 | 0.50 | 0.50 |
| Married | 235 | 0.38 | 0.49 | 605 | 0.41 | 0.49 |
| Has children | 235 | 0.33 | 0.47 | 605 | 0.40 | 0.49 |
| Number of children | 235 | 0.49 | 0.91 | 605 | 0.58 | 0.95 |
| Lives in a rural area | 235 | 0.56 | 0.5 | 605 | 0.55 | 0.50 |
| When of growing age | Obs. | Mean | St. Dev. | Obs. | Mean | St. Dev. |
| Respondent was bedridden for more than one month | 235 | 0.13 | 0.33 | 605 | 0.08 | 0.27 |
| Respondent got malaria | 235 | 0.6 | 0.49 | 605 | 0.55 | 0.5 |
| Respondent got polio | 235 | 0.04 | 0.2 | 605 | 0.02 | 0.15 |
| The characteristics of the respondent's father (or mother's characteristics if respondent never met his father) | Obs. | Mean | St. Dev. | Obs. | Mean | St. Dev. |
| The household asset index of the respondent's father, back when the father was of the respondent's age | 235 | 0.4 | 0.17 | 605 | 0.4 | 0.18 |
| Number of cows owned by the respondent's father, back when the father was of the respondents' age | 154 | 7.53 | 22.03 | 370 | 4.5 | 7.59 |
| The respondent's father has secondary or more, back when the father was of the respondent's age | 174 | 0.57 | 0.5 | 500 | 0.49 | 0.5 |

**Table A.3** Descriptive statistics of children of victims and non-victims of electoral violence.

| | Children of parents who experienced electoral violence during their prenatal, childhood or adolescence | | | Children of parents who did not experience electoral violence during their prenatal, childhood, or adolescence | | |
|---|---|---|---|---|---|---|
| | Obs. | Mean | St. Dev. | Obs. | Mean | St. Dev. |
| Height-for-age z-score of girls | 23 | -0.55 | 1.6 | 51 | -1.53 | 1.47 |
| Height-for-age z-score of boys | 13 | -1.42 | 1.67 | 48 | -0.87 | 1.74 |
| Male | 98 | 0.44 | 0.5 | 316 | 0.54 | 0.5 |
| Child was bedridden for more than one month | 98 | 0.12 | 0.33 | 318 | 0.06 | 0.23 |
| Educational attainment | | | | | | |
|   No formal education | 6 | 1.00 | 1.10 | 23 | 1.70 | 1.55 |
|   Nursery | 30 | 3.20 | 2.41 | 72 | 3.11 | 2.34 |
|   ST1 | 9 | 6.11 | 1.54 | 20 | 5.90 | 1.25 |
|   ST2 | 4 | 7.25 | 0.96 | 19 | 7.74 | 2.28 |
|   ST3 | 2 | 7.00 | 1.41 | 27 | 9.22 | 2.61 |
|   ST4 | 5 | 10.80 | 4.09 | 16 | 8.63 | 2.70 |
|   ST5 | 3 | 9.67 | 2.08 | 12 | 11.17 | 1.27 |
|   ST6 | 9 | 11.33 | 1.80 | 12 | 11.17 | 2.79 |
|   ST7 | 2 | 13.00 | 1.41 | 12 | 9.42 | 5.38 |
|   ST8 | 4 | 12.50 | 2.52 | 7 | 13.71 | 1.38 |
|   Secondary | 4 | 15.50 | 1.73 | 11 | 14.27 | 2.90 |
|   No information provided | 20 | 4.95 | 4.94 | 85 | 4.58 | 3.94 |
| Characteristics of the child's parent (our adult respondents) | Obs. | Mean | St. Dev. | Obs. | Mean | St. Dev. |
| Number of cows household owns at time of the interview | 92 | 1.05 | 1.73 | 281 | 1.29 | 3.14 |
| Parent employment status | 98 | 0.46 | 0.5 | 318 | 0.5 | 0.5 |
| Parent has secondary or more | 93 | 4.66 | 1.69 | 310 | 4.49 | 1.79 |
| Parents' household asset index | 98 | 0.4 | 0.17 | 318 | 0.41 | 0.18 |
| Reside in a rural area | 98 | 0.65 | 0.48 | 318 | 0.59 | 0.49 |

**Table A.4** Placebo tests. ATT of children whose parents were affected by electoral violence during their adulthood only. Coarsened matching results.

|  | Children of parents who experienced electoral violence during their prenatal, childhood or adolescence | | | Children of parents who did not experience electoral violence during their prenatal, childhood, or adolescence | | |
|---|---|---|---|---|---|---|
|  | Obs. | Mean | St. Dev. | Obs. | Mean | St. Dev. |
| Height-for-age z-score of girls | 23 | -0.55 | 1.6 | 51 | -1.53 | 1.47 |
| Height-for-age z-score of boys | 13 | -1.42 | 1.67 | 48 | -0.87 | 1.74 |
| Male | 98 | 0.44 | 0.5 | 316 | 0.54 | 0.5 |
| Child was bedridden for more than one month | 98 | 0.12 | 0.33 | 318 | 0.06 | 0.23 |
| Educational attainment |  |  |  |  |  |  |
| No formal education | 6 | 1.00 | 1.10 | 23 | 1.70 | 1.55 |
| Nursery | 30 | 3.20 | 2.41 | 72 | 3.11 | 2.34 |
| ST1 | 9 | 6.11 | 1.54 | 20 | 5.90 | 1.25 |
| ST2 | 4 | 7.25 | 0.96 | 19 | 7.74 | 2.28 |
| ST3 | 2 | 7.00 | 1.41 | 27 | 9.22 | 2.61 |
| ST4 | 5 | 10.80 | 4.09 | 16 | 8.63 | 2.70 |
| ST5 | 3 | 9.67 | 2.08 | 12 | 11.17 | 1.27 |
| ST6 | 9 | 11.33 | 1.80 | 12 | 11.17 | 2.79 |
| ST7 | 2 | 13.00 | 1.41 | 12 | 9.42 | 5.38 |
| ST8 | 4 | 12.50 | 2.52 | 7 | 13.71 | 1.38 |
| Secondary | 4 | 15.50 | 1.73 | 11 | 14.27 | 2.90 |
| No information provided | 20 | 4.95 | 4.94 | 85 | 4.58 | 3.94 |
| Characteristics of the child's parent (our adult respondents) | Obs. | Mean | St. Dev. | Obs. | Mean | St. Dev. |
| Number of cows household owns at time of the interview | 92 | 1.05 | 1.73 | 281 | 1.29 | 3.14 |
| Parent employment status | 98 | 0.46 | 0.5 | 318 | 0.5 | 0.5 |
| Parent has secondary or more | 93 | 4.66 | 1.69 | 310 | 4.49 | 1.79 |
| Parents' household asset index | 98 | 0.4 | 0.17 | 318 | 0.41 | 0.18 |
| Reside in a rural area | 98 | 0.65 | 0.48 | 318 | 0.59 | 0.49 |

Children are matched based on their ethnicity, sex, and the wealth index of their parents. The regressions are weighted by the resulting coarsened weights. The regressions control for children's characteristics including their sex, ethnicity, province of residence, cohort of birth, and whether have been bedridden for more than one month. Other controls include whether the parent of the child is currently employed, educational attainment, height, cohort of birth, and wealth index. In parentheses are the robust standard errors clustered at the child's ethnicity, birth year, and household levels. Significance levels: ***p < 0.01, **p < 0.05, *p < 0.1.